\documentclass[namedreferences]{SolarPhysics}
\usepackage[optionalrh]{spr-sola-addons} 
\usepackage{graphicx}        
\usepackage{color}           
\usepackage{cases}

\usepackage{url}             

\newcommand{\adv}{    {\it Adv. Spa. Res. }}

\newcommand{\aap}{    {\it Astron. Astrophys. }}
\newcommand{\aaps}{   {\it Astron. Astrophys. Suppl. }}

\newcommand{\apj}{    {\it Astrophys. J. }}

\newcommand{\solphys}{{\it Solar Phys. }}

\begin{document}
\begin{article}
\begin{opening}
\title{Extraction of Active Regions and Coronal Holes from EUV Images Using the Unsupervised Segmentation Method in the Bayesian Framework}

\author{S.~ \surname{Arish}$^{1}$\sep
        M.~\surname{Javaherian}$^{2}$\sep
        H.~\surname{Safari}$^{2}$
         A.~\surname{Amiri}$^{1}$}
\runningauthor{Arish \textit{et al.}}
\runningtitle{%
A Robust Unsupervised Segmentation Method \ldots}
   \institute{$^{1}$ Department of Computer Engineering, University of
Zanjan, University Blvd., 45371-38791, Zanjan, I. R. Iran\\
$^{2}$  Department of Physics, University of Zanjan, University Blvd., 45371-38791, Zanjan, I. R.
                    Iran,~email: \url{m_javaherian@znu.ac.ir}}
\begin{abstract}
The solar corona is the origin of very dynamic events that are mostly produced in active regions (AR) and coronal holes (CH). The exact location of these large-scale features can be determined by applying image-processing
approaches to extreme-ultraviolet (EUV) data.

We here investigate the problem of segmentation of solar EUV images into ARs, CHs, and quiet-Sun (QS) images in a firm Bayesian way. On the basis of Bayes' rule, we need to obtain both prior and likelihood models. To find the prior model of an image, we used a Potts model in non-local mode. To construct the likelihood model, we combined a mixture of a Markov-Gauss model and non-local means. After estimating labels and hyperparameters with the Gibbs estimator, cellular learning automata were employed to determine the label of each pixel.

We applied ~the ~proposed ~method ~to ~a \textit{Solar ~Dynamics ~Observatory/ Atmospheric Imaging Assembly} (SDO/AIA) dataset recorded during 2011 and found that the mean value of the filling factor of ARs is 0.032 and 0.057 for CHs. The power-law exponents of the size distribution of ARs and CHs were obtained to be -1.597 and -1.508, respectively, with the maximum likelihood estimator method. When we compare the filling factors of our method with a manual selection approach and the SPoCA algorithm, they are highly compatible.
\end{abstract}
\end{opening}
\section{Introduction}
Studying coronal features and their time-varying properties may allow us to investigate
the relations between events (flares, coronal mass ejections (CMEs), jets, bright points, etc.)
and overall solar activity (Teske and Thomas, 1969; Tan \textit{et al.},
2010; Kilcik \textit{et al.}, 2011).

The solar corona is characterized by a three-part structure consisting of active regions (AR), the quiet Sun (QS), and coronal holes (CH). When the Sun is observed in soft X-rays and extreme ultraviolet (EUV) wavelengths, very bright areas cover a small fraction of the full disk. These areas are called active regions. Numerous dynamic events such as flares, jets, and plasma heating occur within these areas, resulting in the release of enormous quantities of
energy (Aschwanden, 2006; Priest, 2014).

In addition to ARs, dark areas mostly emerge near the northern and
southern polar zones of the Sun with plasma of lower density and an open configuration of magnetic
field lines. These are termed coronal holes \cite{Altschuler}. Because the plasma
within these areas is most of the time depleted, CHs appear much darker than the other
parts of the Sun (AR and QS). Here, we mainly focus on segmentations of
ARs and CHs, and QS would be attainable as the complement of the identified regions.

The increasing amount of solar data necessitates automated methods
for the detection of solar features. This need has dramatically increased in recent years.
To address this need, many types of automatic segmentations and identifications of
solar features have been proposed for observations obtained with solar instruments (\textit{e.g.}, \textit{Solar Dynamics Observatory}, \textit{Transition Region and Coronal Explorer}, and \textit{Big Bear Solar Observatory}).

A sample of methods based on region-growing techniques employed for extracting
ARs in magnetogram images includes boundary-extraction \cite{McAteer}, which distinguishes between
consecutive images to remove both quiet-Sun and transient magnetic features \cite{Higgins}. It also uses local intensity thresholding, median filtering and morphological operations
at 195 \AA ~and magnetograms \cite{Benkhalil}. Another method was proposed by
Kestener \textit{et al.} (2010) on the basis of the wavelet transform
modulus maxima method. In new efforts for visual segmentation, pattern-recognition
methods (both supervised and unsupervised techniques) were developed to
increase accuracy and stability. Some supervised techniques were established based
on Bayesian classifiers \cite{Wit} and fuzzy rules \cite{Colak}. On the other hand,
the unsupervised techniques provide a focus for characterizing coronal regions
consisting of a method that is based on optimization and accurate definition of
clusterings in the regions of interest \cite{Barra2}.

Another significant and perhaps the most efficient classification method to
derive distinguishing properties of coronal features is the spatial possibilistic
clustering algorithm (SPoCA) \cite{Delouille}, which has been developed mainly
to determine the boundaries of ARs and CHs. This algorithm uses the Fuzzy
C-Means algorithm (FCM), and a corrected version of the Possibilistic C-means
algorithm (PCM) to integrate neighboring intensity values. For detailed
information about coronal image processing and visualization and comparison between
both the segmentations and overall detection performances, we refer to Ireland and
Young (2009), Revathy \textit{et al.} (2005), and Verbeeck \textit{et al.} (2013), respectively.

In image-processing categories, the segmentation algorithms are generally expanded
into three groups: a first group of algorithms employs region-based methods,
edge-based methods are approaches used for determining boundaries,
and hybrid approaches are methods that allow us to use the two algorithms mentioned
above.

Our ~proposed region-based ~segmentation ~algorithm is ~designed based on the ~Bayesian framework.
To do this, a posterior model is designated to labels, and in this way, we need to obtain both prior
and likelihood models rooted in Bayes' rule. The Potts model \cite{Wu} is considered as
the prior model of labels \cite{Deng}. In addition, a Markov-Gauss model (MGM) and non-local means (NL-means)
are combined to achieve a likelihood model \cite{Bresson}. The idea of using these new models, which was employed
in the field of noise-reduction models for the first time (Gilboa and Osher, 2007),
is an efficient method among other
recent standard developments (Dong \textit{et al.}, 2011; Werlberger \textit{et al.}, 2010;
Gilboa and Osher, 2007; Takeda \textit{et al.}, 2009). The Gibbs estimator is
employed to estimate labels and hyperparameters \cite{Geman}. The procedure
of cellular learning automata is added to the process to sample labels \cite{Beigy}.

Here, we checked our code on solar data obtained from
EUV full-disk images (171 \AA ~and 193 \AA) recorded by
SDO/AIA to extract ARs and CHs.

The article is organized as follows: we describe the data in
Section \ref{Data}, explain the segmentation method applied on EUV images in the Bayesian framework
in Section \ref{Segmentation}, and present the results and a discussions in Section \ref{Res}.
Finally, the results are interpreted in Section \ref{Conc}.
\section{Description of data sets}\label{Data}
The modern space telescope \textit{Solar Dynamics Observatory} (SDO) was launched on
February 11, 2010 to study the solar atmosphere in many wavelengths with high
temporal cadence and high spatial resolution. The SDO with the instrument
\textit{Atmospheric Imaging Assembly} (AIA: Lemen \textit{et al.}, 2012) provides full-disk imaging of
the Sun in several UV and EUV bands with a pixel size of about 0.6$^{\prime\prime}$.
Here, SDO/AIA 171 \AA ~and 193 \AA ~data sets are employed to test our segmentation code.
As the evolution of ARs and explosive events are best observed
 at 171 \AA ~(Innes and Teriaca, 2013; Schmieder \textit{et al.}, 2013) and
CHs are well-detectable features at 193 \AA ~\cite{Reiss}, our code is prepared to detect ARs and CHs at 171 and 193 \AA, respectively. The normalization of image intensities \cite{Barra1} and modification of limb brightening \cite{Verbeeck} are applied on the dataset as preprocessing steps.

To study the physical properties of ARs and CHs (\textit{e.g.}, filling
factors and brightness ~fluctuations), our ~automatic recognition ~code ~was applied
to an SDO$/$AIA 171 and 193 \AA ~dataset recorded during 2011, from 1 January to 31 December,
taken at 13:00 UT at a cadence of one image per day.
\section{The Segmentation Method}\label{Segmentation}
The following statistical concepts are briefly discussed. The segmentation model based on Bayes' rule was employed to identify pixel labels. In the Markovian process, the intensity of each pixel (each state) only depends on the four nearest neighbors (the previous state) \cite{Brault}, named the mixture of Markov--Gauss model (MGM) \cite{Ayasso}. In the present work, the spatial dependence of pixels was considered nonlocally (the higher order of neighborhoods) Markovian. To do this, the pixels were weighted based on nonlocal means (NL-means). Next, an image was defined as a lattice-like system, wherein intensity and other properties of pixels such as orientation of edges were the same as states of atoms. By assigning an energy function to the image, Gibbs distributions were then computable \cite{Geman}. To model the image, a Markov random field (MRF) was used, which is a set of random variables consisting of a Markov property dissected by an undirected graph \cite{Kassaye}. We can explain images by an assembly of nodes corresponding to pixels. The Potts model represents models for interacting spins (equivalent to pixels) on a crystalline lattice (equivalent to the image) by a simple Hamiltonian \cite{Lucchi}. Lastly, cellular learning automata (CLA), a mixture of cellular automata (CA) algorithm and learning automata (LA) method, were employed to estimate certain pixel labels \cite{Beigy}.

\subsection{The Segmentation Model}\label{The segmentation model}
Let ${\bf X}=(x_1,x_2, ...,x_N)$ be the intensity of pixels in the segmentation
modeling of an image. It is assumed that the image ${\bf X}$ is the union of $K$ disjoint homogeneous
regions (\emph{e.g.,} ARs, CHs, and QS). Each region can be shown with labels represented by ${\bf z}=\{z_{\bf r},{\bf r}\in R \}$, in which each pixel is related to a label $z_{\bf r}\in \{1, ...,K\}$, and R represents the set of all pixels (union of regions).

The selection of labels for regions concerning the whole of the possible states, which is related to the intensity of pixels, can be interpreted as an inference Bayesian problem. Using Bayes' rule, the posterior probability distribution of labels of pixels is given as
\begin{eqnarray}
P({\bf z}|{\bf X},{\bf \theta})\propto P({\bf X}|{\bf z},{\bf \theta}_{\bf X})P({\bf z}|{\bf \theta}_{\bf z}),\label{PosPr}
\end{eqnarray}
where, $P({\bf z}|{\bf X},{\bf \theta})$ is the posterior distribution of labels. The unknown hyperparameters, ${\bf \theta}_{\bf X}$ and ${\bf \theta}_{\bf z}$, are related to the image model (or likelihood model) $P({\bf X}|{\bf z},{\bf \theta}_{\bf X})$ and the labels model (or prior model) $P({\bf z}|{\bf\theta}_{\bf z})$, respectively.
\subsubsection{The Image Model}\label{The image model}
The model $P({\bf X}|{\bf z},{\bf \theta}_{\bf X})$ has the role of the likelihood distribution in Equation (\ref{PosPr}). The image intensity within the i-th region only depends on the corresponding label. For both the independent and identically distributed form, the image model is given as
\begin{eqnarray}
P({\bf X}|{\bf z},{\bf \theta}_{\bf X}) \propto\ \exp\left(-\frac{1}{2}\sum_{k=1}^{K}\sum_{{\bf r}\in R_k}\frac{(X_{\bf r}-m_k)^{2}}{v_k}\right),~~~~~ \theta_X=({\bf m},{\bf v}),\label{iid}
\end{eqnarray}
where $R_k$ is the set of pixels within the $k$-th region. The parameters $m_k$ and $v_k$ are the mean and the variance of the pixel values in the k-th region, respectively. Starting with the pioneering work of Ayasso and Djafari (2010), the MGM is defined by
\begin{eqnarray}
P({\bf X}|{\bf z},{\bf \theta}_{\bf X})\propto\ \exp \left( -\frac{1}{2}\sum_{k=1}^{K}\sum_{{\bf r}\in R_k}\frac{(X_{\bf
r}-\mu_{k}({\bf r}))^{2}}{v_k} \right). \label{MGM}
\end{eqnarray}
The parameter $\mu_{k}({\bf r})$ is defined as follows:
\begin{subnumcases}{\label{lmu} \mu_{k}({\bf r})\equiv}
  m_k,    & {\rm if} $C_{\bf r}=1$, \\
 \frac{1}{|\xi_{\bf r}|}\sum_{{\bf r'}\in \xi_{\bf r}} X_{\bf r'},  & {\rm if} $C_{\bf r}=0$,
\end{subnumcases}
where $\xi_{\bf r}$ is the set of the four nearest neighbors of pixel ${\bf r}$. The parameter $C_{\bf r}=1-\prod_{{\bf r'}\in \xi_{\bf r}} \delta(z_{\bf r'}-z_{\bf r})$ is representative of contours, wherein $\delta(z)$ is the delta Dirac function. For pixels located within contours, the value of $C_{\bf r}$ equals 1; otherwise, it is set to zero.

Each pixel is considered a Markov random variable, where the central pixel  $X_{\bf c}$ is the new state of Markovian process $X_{n+1}$, and neighboring pixels are the previous state of Markov $X_n$. Equation (\ref{lmu}) shows that the mean value of adjacency connectivity is not used to avoid edge smoothing within the contours.

In the present model, the NL-means are used based on nonlocal information of weighted pixels. These weights are computed using the similarity criterion defined by Gestalt-grouping principles \cite{Yoon}. In the nonlocal method, all image pixels can be considered as neighboring pixels.

The nonlocal mean of an image is given by
\begin{eqnarray}\label{NLC}
&&NL(X_{\bf r})=\frac{1}{S_{\bf r}}\sum_{{\bf r'}\in \xi_{\bf r}} \omega_{{\bf r},{\bf r'}}X_{\bf r'},
\end{eqnarray}
where $S_{\bf r}=\sum_{{\bf n}\in \xi_{\bf r}} \omega_{{\bf r},{\bf n}}$ is the normalization factor and $\omega_{{\bf r},{\bf r'}}$ is the weighting factor between the central pixel and the neighboring pixels gained by criteria including brightness similarity and geometric proximity \cite{Yoon}.

By using the mean of nonlocal weights instead of the local mean in Equation (\ref{lmu}), the image model is modified as
\begin{subnumcases}{\label{wmu} \mu_{k}({\bf r})\equiv}
  m_k,    & {\rm if} $C_{\bf r}=1$, \\
 \frac{1}{S_{\bf r}}\sum_{{\bf r'}\in \xi_{\bf r}} \omega_{{\bf r},{\bf r'}}X_{\bf r'},  & {\rm if} $C_{\bf r}=0$.
\end{subnumcases}
Thus, the image intensity model is represented as
\begin{eqnarray}
P({\bf X}|{\bf z},{\bf \theta}_{\bf X})\propto\ \exp \left(-\frac{\alpha}{2}\sum_{k=1}^{K}\sum_{{\bf r}\in R_k}\frac{(X_{\bf
r}-\mu_{k}({\bf r}))^{2}}{v_k}\right), \label{NLMGM}
\end{eqnarray}
where $\alpha$ is the constant that controls the influence of the likelihood model on the output. By increasing $\alpha$, the value of likelihood probability decreases, and \textit{vice versa}.
\subsubsection{Prior Model of Labels}\label{Potts}
We employed an MRF framework to obtain different prior spatial models to model the labels. MRF consists of a set of random variables ${\bf F}=\{F_{\bf r}\in R\}$ such that every one of the random variables $F_{\bf r}$ arises from a discrete set ${\bf L}=\{l_1,...,l_k\}$ \cite{Boykov}. Each random variable $F_{\bf r}$ and its neighborhoods $\xi_{\bf r}$ is considered a clique. The probability distribution of ${\bf F}$ is included in the Gibbs distribution as follows \cite{PETROUDI}:
\begin{eqnarray} \label{PF}
P({\bf F})=\frac{\exp (-U({\bf F}))}{\sum_{{\bf F}} \exp(-U({\bf F}))}.
\end{eqnarray}
The energy function $U$ and the probability value $P({\bf F})$ correspond to different configurations of ${\bf F}$ \cite{Boykov}. The random variables are considered as labels corresponding with regions. Using the Potts model, the energy function is defined as  \cite{Deng}
\begin{eqnarray} \label{Uz}
U({\bf z})= \beta \sum_{{\bf r}\in R}\sum_{{\bf r'}\in \xi_{\bf r}}(1-\delta(z_{\bf r}-z_{\bf r'})),
\end{eqnarray}
where $\beta$ is a coefficient that controls the rate of smoothness (the amount of connections created among labels). Using Equations (\ref{PF}) and (\ref{Uz}), the probability function of the Potts model can be expressed as
\begin{eqnarray}
&&P({\bf z}|\beta)\propto \exp\left(\beta \sum_{{\bf r}\in R}\sum_{{\bf r'}\in \xi_{\bf r}}\delta(z_{\bf r}-z_{\bf r'})\right).
\end{eqnarray}
If the smoothness in segmentation increases, the probability value shows the growth in its distribution. Choosing a greater value for $\beta$ causes loss of segmentation details, and the selection of lower values may lead to increased noise \cite{McGrory}. For nonlocal information, the probability can be expressed as \cite{Werlberger}
\begin{eqnarray}\label{finalbeta}
P({\bf z}|\beta)\propto \exp\left( \sum_{{\bf r}\in R}\sum_{{\bf r'}\in \xi_{\bf r}}\beta \omega_{{\bf r},{\bf r'}}\delta(z_{\bf r}-z_{\bf r'})\right).
\end{eqnarray}
\subsection{The Optimization Model and Parameter Estimation}
The labels of pixels ${\bf \hat z}$ and hyperparameters ${\bf \hat \theta}$ can be obtained by solving the following equation
\begin{eqnarray}\label{JMAP}
\{{{\bf \hat z} },{{\bf \hat \theta}}\}=\arg \max_{{\bf z},{\bf \theta}}(P({\bf z},{\bf \theta}|{\bf X}))~~~~~~~~~~~~~~~~~~~\nonumber \\
=\arg \max_{{\bf z},{\bf \theta}}(P({\bf X}|\bf z,{\bf \theta}_X)P({\bf z}|\theta_{\bf z})P({\bf \theta})),
\end{eqnarray}
where $P({\bf \theta})$ represents the probability model of the set of unknown parameters. Using Gibbs sampling \cite{Geman}, ${\bf \hat z}$ and ${\bf \hat \theta}$ are obtained in the iteration loop as follows
\begin{eqnarray}\label{zmcmc}
{\bf z}(n)\propto P\left({\bf X}|{\bf z}(n-1),\theta_{\bf X}(n-1)\right)P\left({\bf z}(n-1)|\theta_{\bf z}\right),
\end{eqnarray}
\begin{eqnarray}\label{tmcmc}
{\bf \theta}(n)\propto P\left({\bf X}|{\bf z}(n),\theta_{\bf X}(n-1)\right)P\left({\bf \theta}(n-1)\right).
\end{eqnarray}
According to this sampler, the $n$-th sample of ${\bf z}$ is obtained from Equation (\ref{zmcmc}). Then, we use it in Equation (\ref{tmcmc}) to sample $\theta $, and \textit{vice versa}. This iterative way must be continued until
\begin{eqnarray}\label{epsilon}
||z_{\bf r}(n)- z_{\bf r}(n-1)|| < \varepsilon.
\end{eqnarray}
Parameter $\varepsilon$ is related to the precision of segmentation. For more details about estimating all the unknown parameters we refer to Humblot and Djafari (2006), and Murphy (2007).
\subsubsection{Label Estimation}
The cellular learning automata method, a mathematical model for complex dynamical systems consisting of numerous simple elements, was exploited to estimate labels \cite{Beigy}.

Each cell sends an output (action) $\eta_i\in \{\eta_1,...,\eta_K\}$ to its surroundings and then receives an input response $\phi\in\{0,1\}$. The probability of choosing each action in the $n$-th iteration is expressed by the action probability ${\bf p}(n)\in \{p_1(n),...,p_K(n)\}$. In other words, when the cell receives a desired response input $\phi=1$, the action for automaton $i$, which is denoted by $\eta_i$, is rewarded; for another possible input $\phi=0$, it is penalized \cite{Narendra}. In CLA, input $\phi$ is determined by the current output (action) of neighbors of each cell. Here, pixels are equivalent to cells, and actions are comparable with labels. For the label of each pixel, all probabilities of the number of actions applied to each pixel are calculated. In other words, the greater the value of action probability of each label, the higher the selection probability of the label. The label of each pixel in each iteration is attained by
\begin{eqnarray}
z_{\bf r}(n)={\rm IndMax}_{k=1,...,K}(\widehat{r}_{k}.p_{k}(n-1)),
\end{eqnarray}
where the function IndMax returns the index corresponding with the highest value that
can be interpreted as the label of the pixel, and $\widehat{r}_{k}$ is a random number
ranging from 0 to 1 assigned to $k$-th selection probability.

The Potts model can be regarded as a CA that is able to specify the posterior probability of labels $P(z_{\bf r}|{\bf X},{\bf \theta})$ for each cell based on two criteria: Markov dependence (image model) and similarity between each cell and its neighbors (labels model). On the basis of the mentioned assumptions, we must consider the desired input response (label), $\phi=1$ obtained from the action increasing the posterior probability; on the other hand, $\phi=0$ reduces the probability \cite{Narendra}.

In $n$-th iteration, the probability related with the value of the label $z_{\bf r}$ is reinforced. Thus, the probability related to $z_{\bf r}$ is reinforced in matrix ${\bf Pm}$ (a $N\times K$ matrix in which the element $(r,k)$ shows the assignment probability of the $k$-th label to the $r$-th pixel). By computing ${\bf Pm}$, the final label, $Z_{o_{\bf r}}$ is given as
\begin{eqnarray}
Z_{o_{\bf r}}={\rm IndMax}_{k=1,\ldots,K}({\bf Pm}_k).
\end{eqnarray}
\section{Results and Discussion}\label{Res}
\begin{figure}
\centerline{\includegraphics[width=2.5\textwidth,clip=]{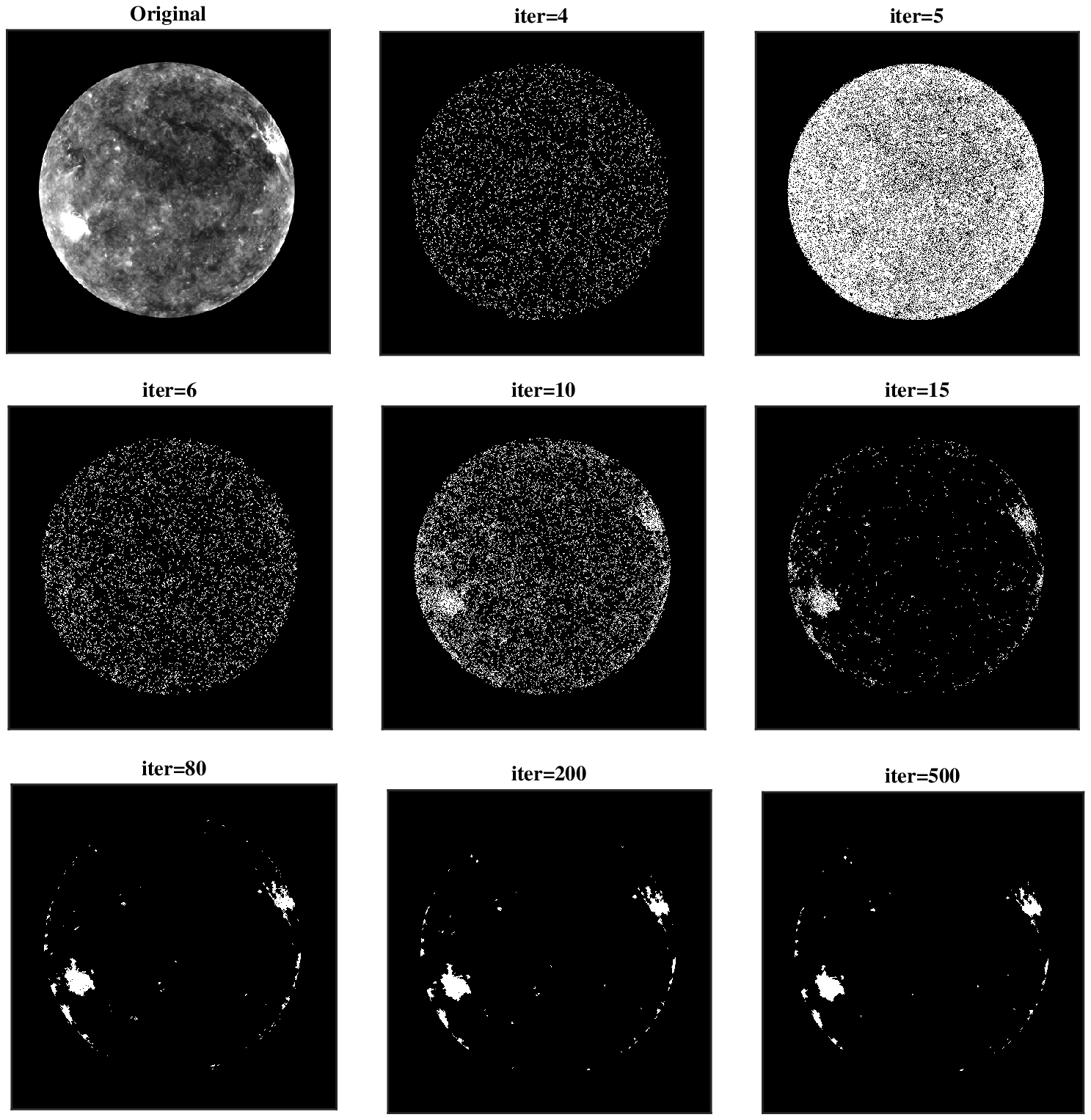}}
\caption[]{The SDO$/$AIA data recorded on 27 November 2010 (08:23:13 UT) at 171 \AA ~(first image) with different numbers of iterations. By increasing the number of iterations in specified $\alpha$ and $\beta$, ARs are gradually recognized. After about 400 iterations, the filling factors of ARs do not change significantly.} \label{fig2}
\end{figure}

The proposed method was applied to solar data that have mainly been recorded at two wavelengths (171 and 193 \AA).
It is a well-established fact that ARs can be identified better at 171 \AA, and on the other hand, the preferred wavelength for CH recognition is 193 \AA. To improve image segmentations, we here extracted the optimized values for parameters $\alpha$ and $\beta$ specified in Equations (\ref{NLMGM}) and (\ref{finalbeta}), respectively. We determined regions in one computational loop with $n$ iterations with the condition given in Equation (\ref{epsilon}). Here, we decided that the parameter $\varepsilon$ equals 30 pixels for images of size 1024 $\times$ 1024 pixels to obtain great precision. Figure \ref{fig2} shows different numbers of iterations applied to 171 \AA ~data to segment ARs. It can be seen that ARs are gradually recognized in higher-order iteration. Certain performances of both AR and CH identification reach their final value in the 400-th and 1200-th iteration, respectively.

\begin{figure}
\centerline{\includegraphics[width=1.45\textwidth,clip=]{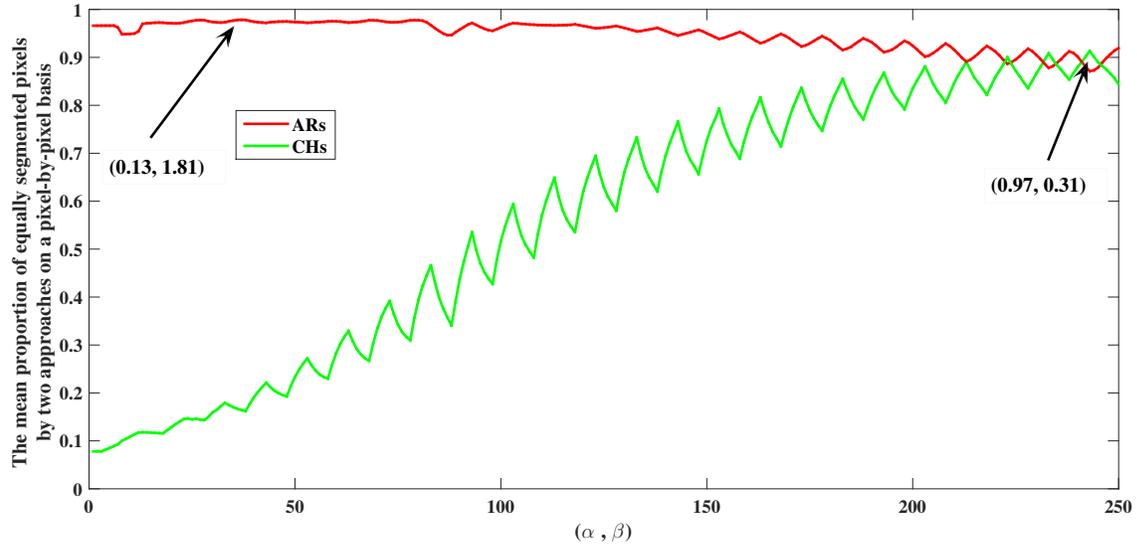}}
\caption[]{The two segmented images (manual selection approach and proposed method) are compared on a pixel-by-pixel basis, measuring the proportion of equally segmented pixels. For example, if a pixel is segmented as AR in both results, it is assigned a score (= 1), if a pixel is segmented as non-AR in both results, it is assigned a score (= 1), if a pixel is segmented as AR in one result and as non-AR in the other result, it is assigned a score (= 0). Then, the summation of scores is divided by the number of pixels of the disk. The mean value of the summations for 100 images taken at 171 \AA ~(wavelength for AR recognition) and 193 \AA ~(wavelength for CH recognition) for different values of $\alpha$ and $\beta$ are plotted with red and green lines, respectively. When the corresponding value approaches unity, similar regions in both our method and the manually extracted approach increase. The set ($\alpha$, $\beta$) varies in the form of (0.01,0.01), (0.01,0.31) \ldots (0.05,0.01), (0.05,0.31) \ldots (0.97,2.41), (0.97,2.71). The parameters  $\alpha$ and $\beta$ range from 0 to 1, and 0 to 3, respectively. For best results, the set ($\alpha$, $\beta$) is approximately equal to (0.13, 1.81) and (0.97, 0.31) for ARs and CHs, respectively. We indicate the optimal ($\alpha$, $\beta$) combinations for AR and CH with arrows.} \label{fig3}
\end{figure}

In the next step, 200 images recorded (from 2010 to 2013) at two wavelengths were randomly selected. ARs and CHs were manually extracted from 171 and 193 \AA ~images, respectively. The code with different values of $\alpha$ and $\beta$ was applied to these data to distinguish regions.
The two segmented images (manual selection approach and proposed method) were compared on a pixel-by-pixel basis, measuring the proportion of equally segmented pixels. For example, if a pixel is segmented as an AR in both results, it is assigned a score (= 1); if a pixel is segmented as non-AR in both results, it is assigned a score (= 1); if a pixel is segmented as AR in one result and as non-AR in the other result, it is assigned a score (= 0). Then, the summation of scores is divided by the number of pixels of the disk. Figure \ref{fig3} demonstrates the mean value of the summations for 100 images at each wavelength for different values of $\alpha$ and $\beta$. Since this segmentation method is the result of a random Markovian process, some negligible changes appeared in the resulting segmented images in each running code. The optimal values for the set ($\alpha$, $\beta$) are obtained to equal (0.13, 1.81) and (0.97, 0.31) for ARs and CHs, respectively.

\begin{figure}
\centerline{\includegraphics[width=2\textwidth,clip=]{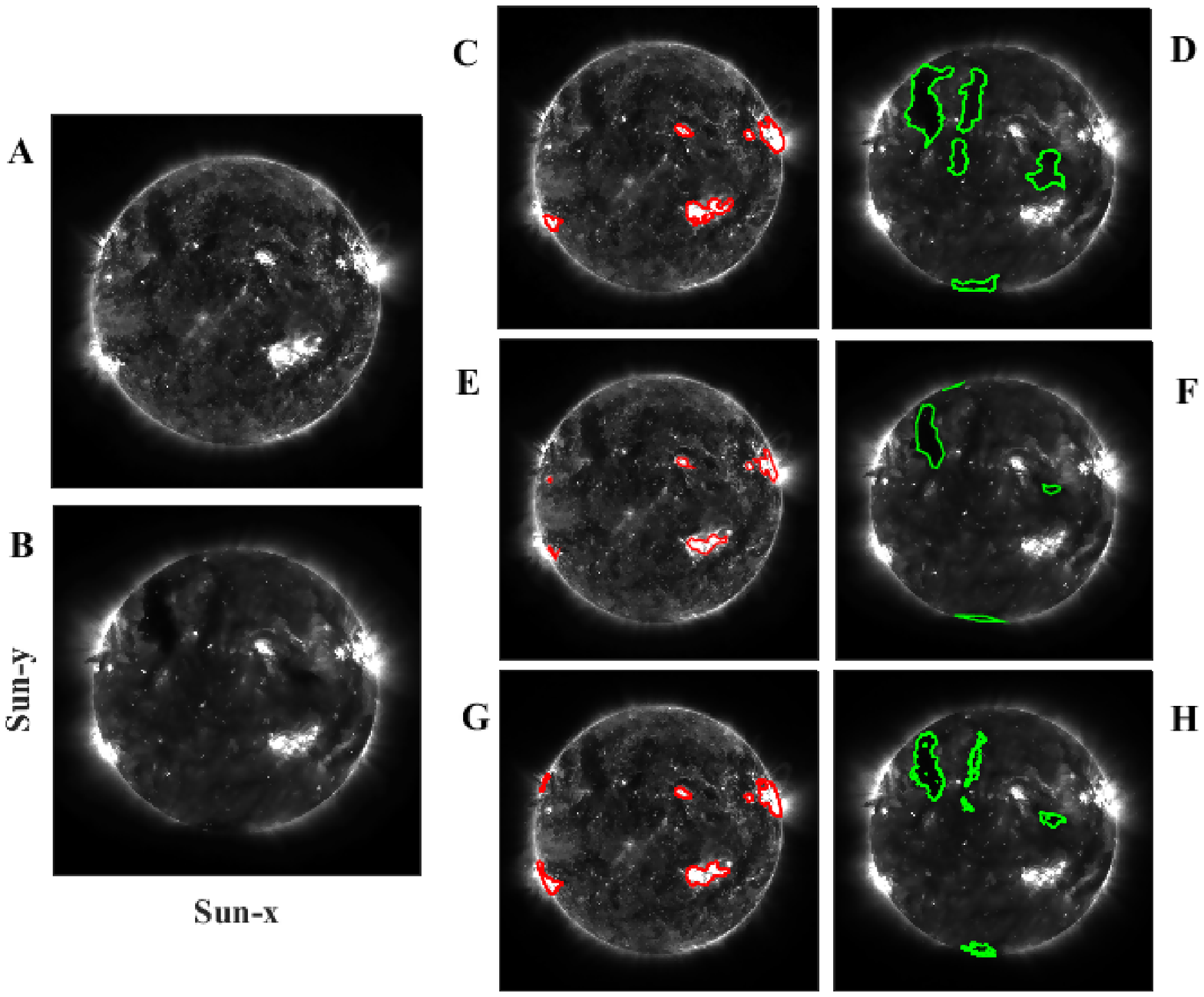}}
\caption[]{The full-disk SDO$/$AIA 171 \AA ~(A) and 193 \AA ~(B) images recorded on 25 June 2011 (13:00:36 UT). In the first row (right panel), the results of the manually extracted approach for the AR map (red contours) (C) and CH map (green contours) (D) are presented. In the second row, the SPoCA results for ARs (red contours) (E) and CHs (green contours) (F) are demonstrated. The results of the present method for identifying ARs (red contours) (G) and CHs (green contours) (H) for the optimal values of ($\alpha$, $\beta$) are presented in the third row.} \label{fig4}
\end{figure}

\begin{figure}
\centerline{\includegraphics[width=2.1\textwidth,clip=]{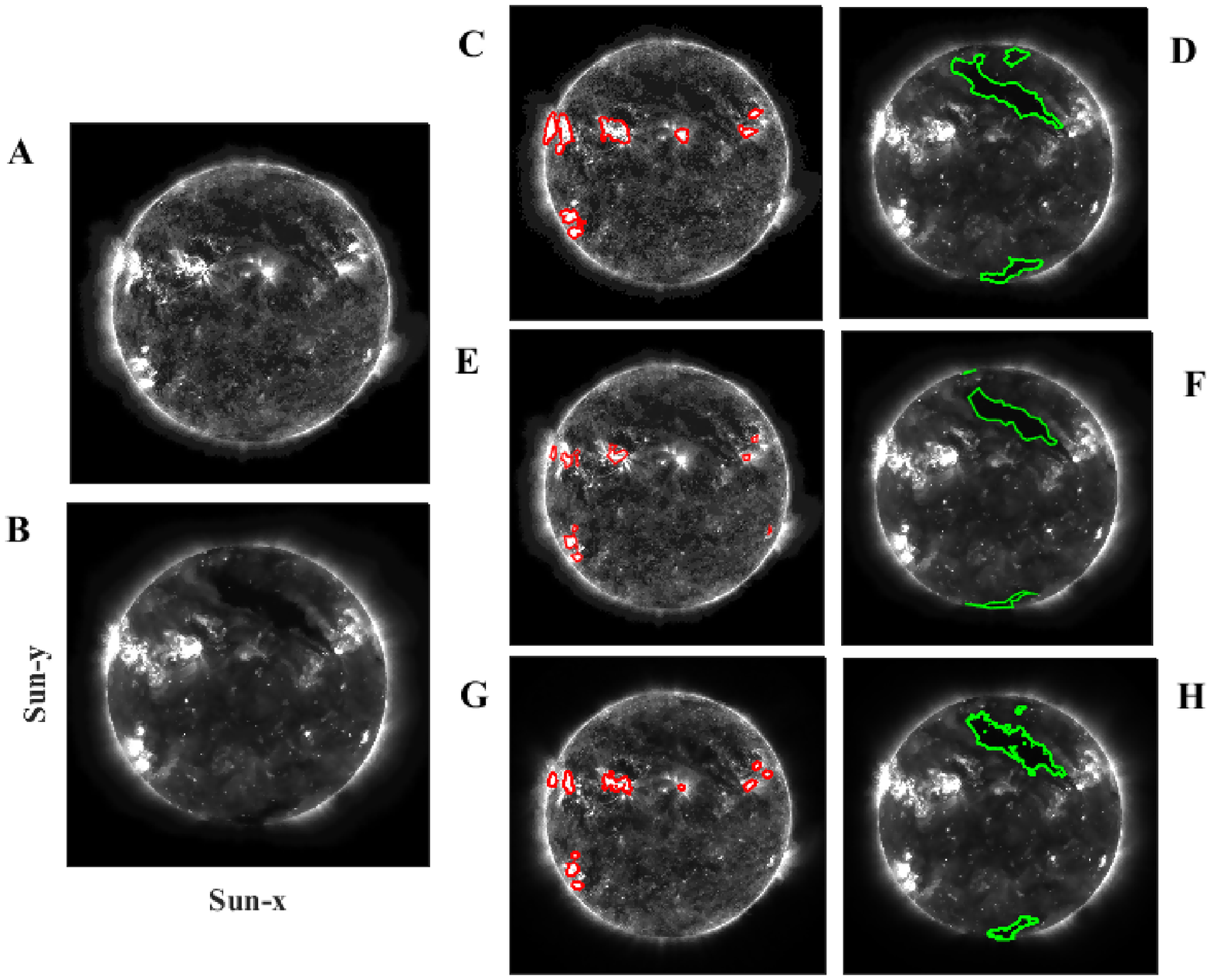}}
\caption[]{The full-disk SDO$/$AIA 171 \AA ~(A) and 193 \AA ~(B) images recorded on 28 July 2011 (13:00:37 UT) are shown. In the first row (right panel), the results of the manually-extracted approach for AR map (red contours) (C) and CH map (green contours) (D) are presented. In the second row, the SPoCA results for ARs (red contours) (E) and CHs (green contours) (F) are demonstrated. The results of the present method for identifying ARs (red contours) (G) and CHs (green contours) (H) for the optimal values of ($\alpha$, $\beta$) are presented in the third row.} \label{fig5}
\end{figure}

To estimate the reliability of the algorithm and to compare it to the SPoCA results, we picked up full-disk SDO/AIA images recorded on 25 June 2011 (13:00:36 UT) and 28 July 2011 (13:00:37 UT) taken at 171 (Figures \ref{fig4}A, and \ref{fig5}A) and 193 \AA ~(Figures \ref{fig4}B, and \ref{fig5}B). First, we manually extracted ARs (Figures \ref{fig4}C, and \ref{fig5}C) and CHs (Figures \ref{fig4}D, and \ref{fig5}D). For comparison, the performance of the SPoCA results in the HEK catalog \cite{Hurlburt} represented at 171 \AA ~(Figures \ref{fig4}E, and \ref{fig5}E) and 193 \AA ~(Figures \ref{fig4}F, and \ref{fig5}F) are shown. Then, our method was applied on these two wavelengths to extract ARs (Figures \ref{fig4}G, and \ref{fig5}G) and CHs (Figures \ref{fig4}H, and \ref{fig5}H) with the optimal values of ($\alpha$, $\beta$).

\begin{figure}
\centerline{\includegraphics[width=3.6\textwidth,clip=]{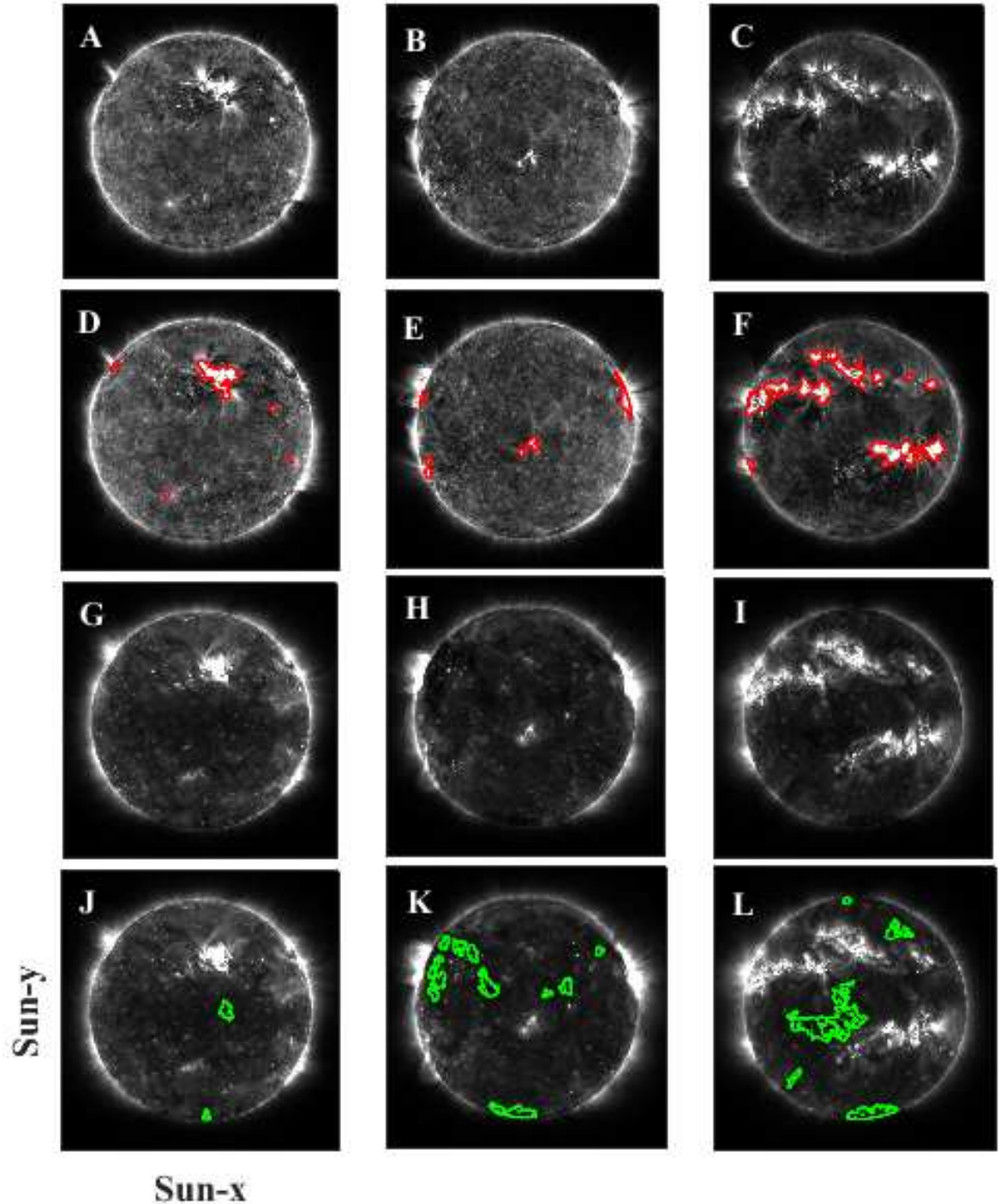}}
\caption[]{Other samples of segmented images are presented. The original images at 171 \AA ~(A, B, and C), and overlay of AR maps (D, E, and F), the original images at 193 \AA ~(G, H, and I), and overlay of CH maps (J, K, and L) are shown. These data are recorded on 22 January 2011 (left column), 24 February 2011 (middle column), and 31 March 2011 (right column), respectively.} \label{fig6}
\end{figure}

To exclude bright features appearing as bright points, very slender coronal loops \cite{Ulmschneider} resolving out of ARs, etc., the minimum sizes for ARs and CHs were considered to be 1400 and 3000 arcsec$^2$, respectively, close to the values in Verbeeck \textit{et al.} (2014). The method clearly is able to separate regions of interest with high precision. Figure \ref{fig6} presents more samples at 171 \AA ~(first row) and 193 \AA~(third row). The recognition of ARs and CHs are displayed with red (second row) and green contours (last row), respectively. Columns are representative of data recorded on 22 January 2011 (left column), 24 February 2011 (middle column), and 31 March 2011 (right column).

\begin{figure}
\centerline{\includegraphics[width=1.4\textwidth,clip=]{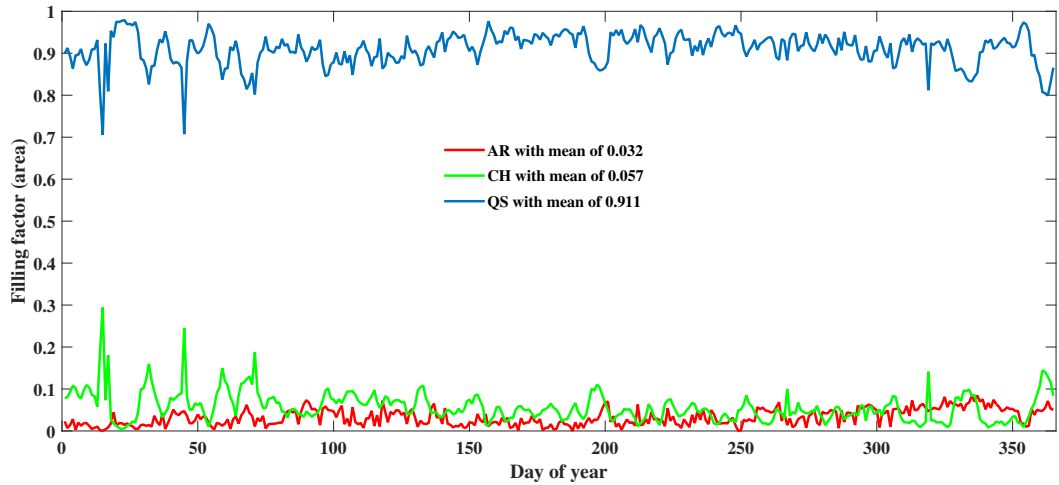}}
\caption[]{Filling factors of AR, CH, and QS obtained with our segmentation code applied on
AIA 171 and 193 \AA ~images from 1 January to 31 December 2011.} \label{fig7}
\end{figure}

\begin{figure}
\centerline{\includegraphics[width=1.4\textwidth,clip=]{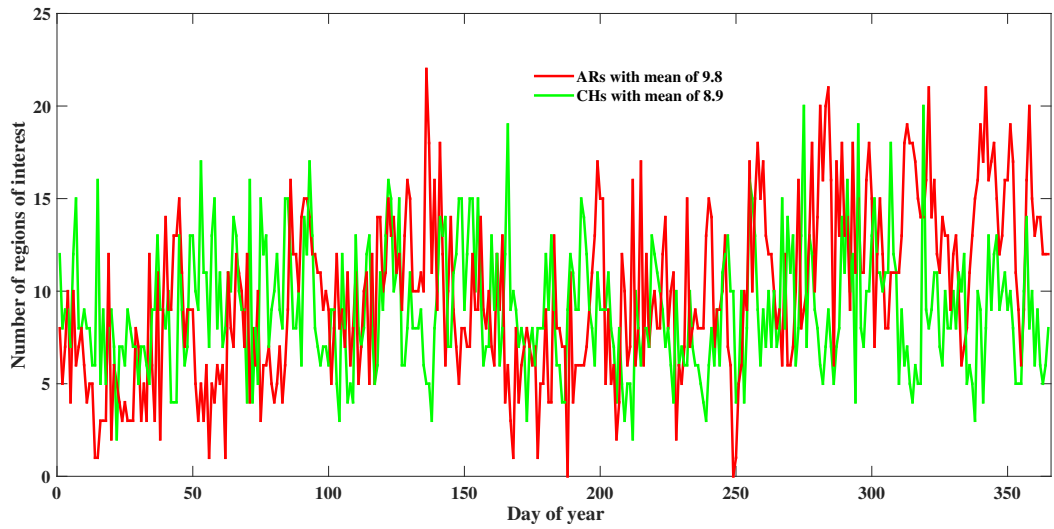}}
\caption[]{The number of daily ARs and CHs obtained from 1 January to 31 December 2011.} \label{fig8}
\end{figure}

\begin{figure}
\centerline{\includegraphics[width=1.35\textwidth,clip=]{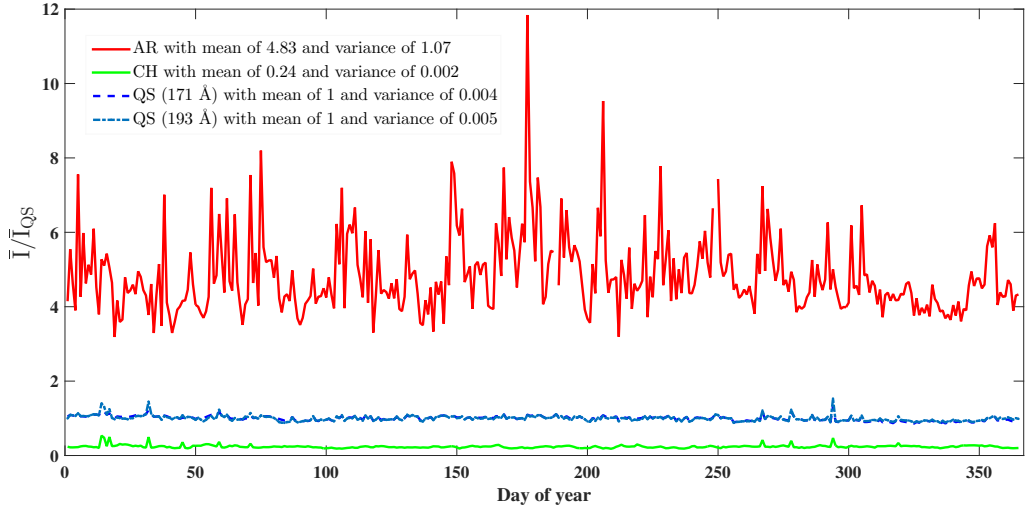}}
\caption[]{Brightness fluctuations of ARs, CHs, and QS with their variances during 2011. We divided the intensities in this plot by the same all-year average value of QS brightness. The all-year average value of QS brightness for 171 and 193 \AA ~we obtained 164 and 155 DN$/$s, respectively. The mean value of brightness of ARs is about 20 times and five times as much as that calculated for CHs and QS, respectively.} \label{fig9}
\end{figure}

To investigate the statistics of coronal regions, the code was applied to a dataset of 12 months ranging from 1 January to 31 December 2011. At 171 and 193 \AA~ an image for each day at each wavelength was selected. The filling factors of AR, CH, and QS are shown with mean values of 0.032, 0.057, and 0.911, respectively (Figure \ref{fig7}). The fluctuations (the solar rotation of about 27 days) are compatible with corresponding results obtained by Verbeeck \textit{et al.} (2014) for 2011. The total number of ARs and CHs extracted by our code for 2011 equal 3580 and 3275, respectively. The number of ARs and CHs on each day with the mean value of 9.8 and 8.9, respectively, are shown in Figure \ref{fig8}. The correlation (Pearson) between these two time series is about zero. Brightness fluctuations of ARs, CHs, and QS with their variances during 2011 are illustrated in Figure \ref{fig9}. The mean brightness value of ARs is 20 and five times as much as that calculated for CHs and QS, respectively. The variance in the brightness of AR is about 200 times more than that computed for the brightness of QS, which is expected to be indicative of their activity.

\begin{figure}
\centerline{\includegraphics[width=1.4\textwidth,clip=]{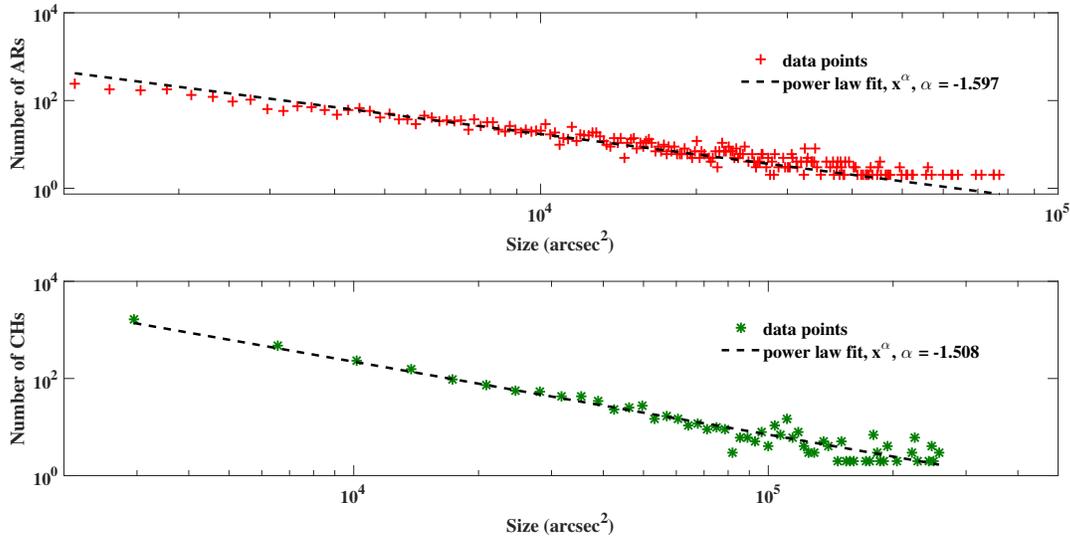}}
\caption[]{Size--frequency distribution of ARs (upper panel) and CHs (lower panel) in $\log$$–-$$\log$ scale. The linear fit (indicated by the dashed line) is drawn based on parameters extracted by the MLE method. The power-law exponents are equal to -1.597 and -1.508 for ARs and CHs, respectively.} \label{fig10}
\end{figure}

The size--frequency distribution of ARs (Figure \ref{fig10}, upper panel) and CHs (Figure \ref{fig10}, lower panel) are displayed in log-log scale. We used a data-based method to determine the optimal bin number within histograms. This optimization method is based on the minimizing loss function between the histogram and an unknown density function by adopting the mean integrated squared error \cite{Shimazaki}. We used the maximum likelihood estimator method \cite{Clauset} to compute the power-law exponents of size--frequency distributions. The power-law exponents $\alpha$, which is shown as slope of linear fit (dashed line) in Figure \ref{fig10}, are equal to -1.597$\pm$0.025 and -1.508$\pm$0.028 for ARs and CHs, respectively, with 97 $\%$ confidence levels in fitting. The scatter plots of ARs (Figure \ref{fig11}, left panel) and CHs (Figure \ref{fig11}, right panel) with mean values for each bin (0--1800 arcsec$^2$, 1800--3600 arcsec$^2$, \textit{etc.}) are presented. Both small ARs and CHs have excessive scatter in brightness.

To check the influence of noise on the segmentation of data taken on 6 February 2013 (12:08:30 UT) with the optimized values of ($\alpha$, $\beta$), was used the zero-mean Gaussian noise. The signal-to-noise ratio (S$/$N) can be obtained by
\begin{eqnarray}
{\rm SNR}=\frac{\sigma_{_{\rm\small{signal}}}^2}{\sigma_{_{\rm\small{noise}}}^2},
\end{eqnarray}
where $\sigma_{_{\rm\small{signal}}}$ and $\sigma_{_{\rm\small{noise}}}$ are the standard deviations of the pixel intensity and noise, respectively. In a similar manner, S$/$N is formulated in decibels as
\begin{eqnarray}
{\rm S/N}_{\rm dB}=10\log({\rm S/N}).
\end{eqnarray}
To evaluate the segmentation error, we calculated the percentage of pixels that were differently classified in the case of artificial noise, when compared to the segmentation without noise in the same way described above for Figure \ref{fig3}. Thus, the segmentation error for ARs (Figure \ref{fig12}, left panel) and CHs (Figure \ref{fig12}, right panel), which was applied to data recorded on 6 February 2013 in different values of S$/$N was calculated when zero-mean Gaussian noise was added to the data. As expected, the influence of noise on the segmentation of ARs is weaker than that of CHs. We can see that in the lower values of S$/$N, the result of segmentation is unpredictable.

The filling factors of ARs (Figure \ref{fig13}, left panel) and CHs (Figure \ref{fig13}, right panel) within 50 images, which were selected randomly from 2010--2012, were extracted using both a manual selection approach and our automatic method for the optimized values of $(\alpha,\beta)$ noted in Figure \ref{fig3}. By comparing these two results on a pixel-by-pixel basis, we noted some false-positive detections.
\section{Conclusions} \label{Conc}
Here, we presented a hybrid algorithm defined in a Bayesian framework based on the
MGM and Potts model. Then, cellular learning automata were employed to sample labels from their
posterior probability. The proposed algorithm was applied to coronal full-disk SDO/AIA data to subdivide
areas into regions of interest. By applying this code to various types of solar data on different
days, its flexibility and capability in separating regions were tested.

\begin{figure}
\centerline{\includegraphics[width=1.4\textwidth,clip=]{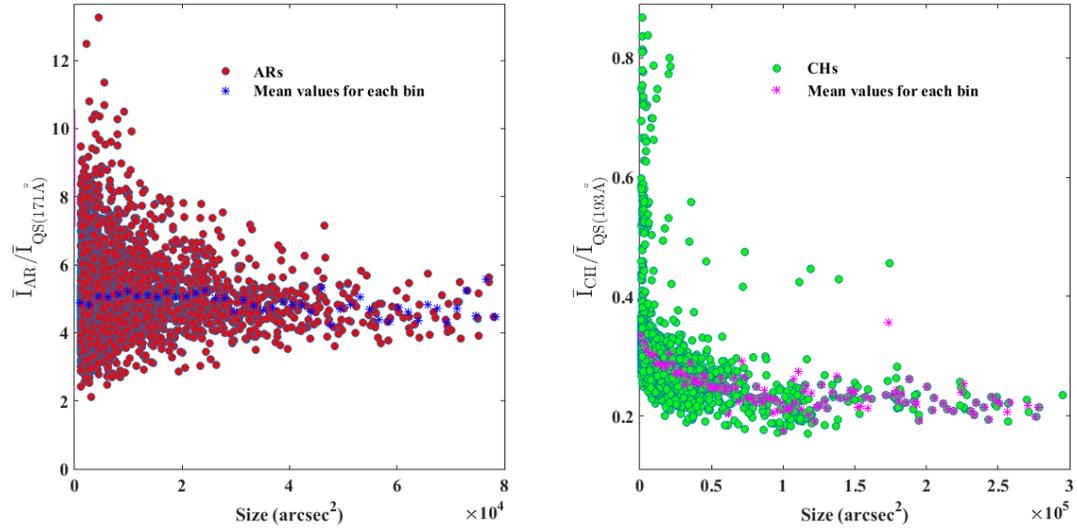}}
\caption[]{Scatter plots of both ARs (left panel) and CHs (right panel) brightness \textit{vs.} size. We divided the intensities in all plots by the same all-year average value of QS brightness. The mean values for each bin (0 -- 1800 arcsec$^2$, 1800 -- 3600 arcsec$^2$, \textit{etc.}) for ARs and CHs are shown as blue and pink stars, respectively. The right panel shows that some small CH regions have high average values in brightness. This suggests that some border pixels of CHs in fact contain a contribution of QS and ARs.} \label{fig11}
\end{figure}

\begin{figure}
\centerline{\includegraphics[width=1.4\textwidth,clip=]{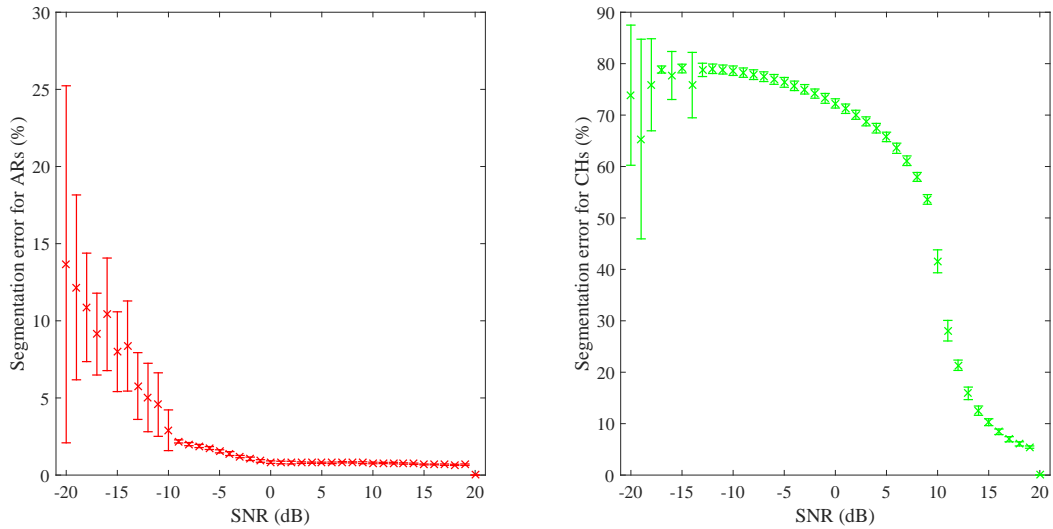}}
\caption[]{The segmentation error \textit{vs.} S$/$R is obtained by adding zero-mean Gaussian noise for ARs and CHs taken at 171 (left panel) and 193 {\AA} (right panel), respectively.} \label{fig12}
\end{figure}

\begin{figure}
\centerline{\includegraphics[width=1.4\textwidth,clip=]{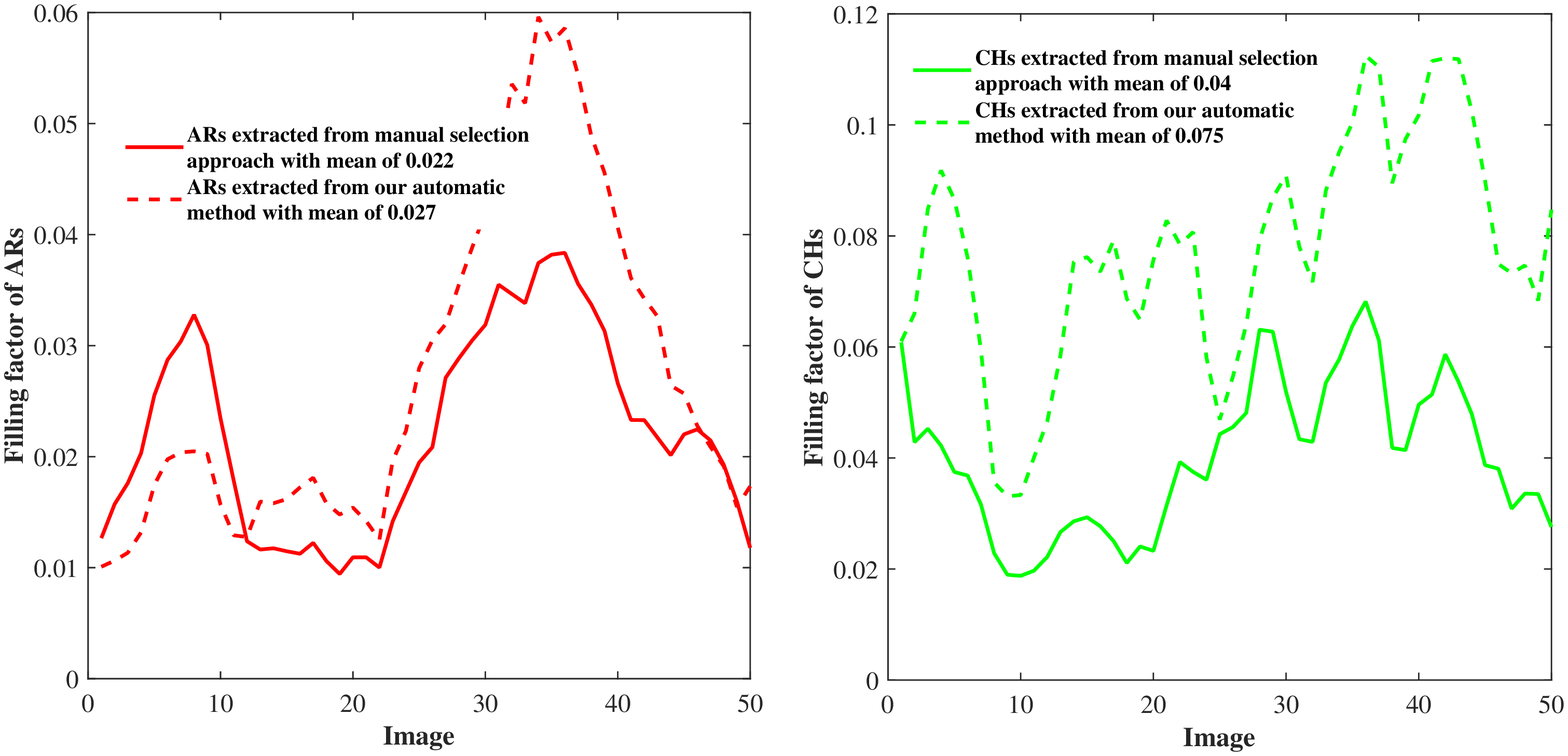}}
\caption[]{The filling factors of ARs (left panel) and CHs (right panel) within 50 images, which were selected randomly from 2010--2012, are extracted using both manual selection approach and our automatic method from the optimal values of $(\alpha, \beta)$, noted in Figure \ref{fig3}. False-positive detections of our code are clearly observable.} \label{fig13}
\end{figure}

In Figure \ref{fig11} we show that CHs with sizes smaller than 5000 arcsec$^2$ are more scattered in brightness, and for large sizes, fluctuations in brightness decrease. For small CH regions, the average brightness increases. This shows that the influence of the border pixels that are located next to the QS and especially in the vicinity of the ARs becomes important. For large CH regions, the average brightness shows less scattering. Most probably, the false-positive detection error of the code (Figure \ref{fig13}, right panel) affects the average value of brightness for some CH regions.

As expected, the segmentation of CHs is more sensitive to noise than the
segmentation of ARs. Therefore, it is necessary to take preprocessing steps before
recognizing CHs. This method can be improved to restore and segment simultaneously
to retrieve coronal regions. It will also be better to revise our
code to use two or three wavelengths to characterize ARs and CHs.

In the future, we will try to establish a reliable criterion for determining border pixels of ARs and CHs, based on mean and standard deviation. Moreover, we intend to revise this algorithm to extract regions of interest from photospheric and$/$or chromospheric data. Then, we aim to investigate the evolution and variations of different types of regions in a number of wavelengths in half a solar cycle.

\begin{acks}
The authors acknowledge the SPoCA group: C. Verbeeck, V. Delouille, and B. Mampaey for making SPoCA results publicly available and the HEK team for providing SPoCA detections in the HEK. The authors also thank the anonymous referee
for meticulous comments and suggestions.
\end{acks}

\newpage


\end{article}

\begin{thebibliography}{}
\bibitem[\protect\citeauthoryear{Altschuler, Trotter, and Orrall}{1972}]{Altschuler}
Altschuler, M.D., Trotter, D.E., Orrall, F.Q.: 1972, \solphys \textbf{26}, 354 (DOI: 10.1007$/$BF00165276).
\bibitem[\protect\citeauthoryear{Aschwanden}{2006}]{Aschwanden1}
Aschwanden, M.J.: 2006, {\it Physics of the Solar Corona}, 2nd edn., Praxis Publishing, Springer (DOI: 10.1007$/$3-540-30766-4).
\bibitem[\protect\citeauthoryear{Ayasso and Djafari}{2010}]{Ayasso}
Ayasso, H., Djafari, A.: 2010, {\it IEEE Tran. Image Process.} \textbf{19}, 2265 (DOI: 10.1109$/$TIP.2010.2047902).
\bibitem[\protect\citeauthoryear{Barra, Delouille, and Hochedez}{2008}]{Barra1}
Barra, V., Delouille, V., Hochedez, J.F.: 2008, \adv \textbf{42}, 917 (DOI: 10.1016$/$j.asr.2007.10.021).
\bibitem[\protect\citeauthoryear{Barra \textit{et al.}}{2009}]{Barra2}
Barra, V., Delouille, V., Kretzschmar, M., Hochedez, J.F.: 2009, \aap \textbf{505}, 361 (DOI: 10.1051$/$0004-6361$/$200811416).
\bibitem[\protect\citeauthoryear{Beigy and Meybodi}{2004}]{Beigy}
Beigy, H., Meybodi, M.R.: 2004, {\it Adv. Comp. Sys.} \textbf{7}, 295 (DOI: 10.1142$/$S0219525904000202).
\bibitem[\protect\citeauthoryear{Benkhalil \textit{et al.}}{2006}]{Benkhalil}
Benkhalil, A., Zharkova, V.V., Zharkov, S., Ipson, S.: 2006, \solphys \textbf{235}, 87 (DOI: 10.1007$/$s11207-006-0023-7).
\bibitem[\protect\citeauthoryear{Li}{2009}]{Boykov}
Li, S.Z.: 2009, {\it Markov Random Field Modeling in Image Analysis}, 3rd edn., Advances in Pattern Recognition, Springer-Verlag London (DOI: 10.1007$/$978-1-84800-279-1).
\bibitem[\protect\citeauthoryear{Brault and Djafari}{2005}]{Brault}
Brault, P., Djafari, A.: 2005, {\it J. Electron. Imaging}, \textbf{14}, 043011 (DOI: 10.1117$/$1.2139967).
\bibitem[\protect\citeauthoryear{Bresson and Chan}{2008}]{Bresson}
Bresson, X., Chan, T.F.: 2008, {\it UCLA. cam. report}, 08.
\bibitem[\protect\citeauthoryear{Clauset, Shalizi, and Newman}{2009}]{Clauset}
Clauset, A., Shalizi, C.R., Newman, M.E.J.: 2009, {\it SIAM Review}, \textbf{51}, 661 (DOI: 10.1137$/$070710111).
\bibitem[\protect\citeauthoryear{Colak and Qahwaji}{2013}]{Colak}
Colak, T., Qahwaji, R.: 2013, \solphys \textbf{283}, 143 (DOI: 10.1007$/$s11207-011-9880-9).
\bibitem[\protect\citeauthoryear{Deng and Clausi}{2004}]{Deng}
Deng, H., Clausi, D.A.: 2004, {\it Pattern Recognition}, \textbf{37}, 2323 (DOI: 10.1016$/$j.patcog.2004.04.015).
\bibitem[\protect\citeauthoryear{Dong \textit{et al.}}{2011}]{Dong}
Dong, W., Zhang, D., Shi, G., Wu, X.: 2011, {\it IEEE Tran. Image Process.} \textbf{20}, 1838 (DOI: 10.1109$/$TIP.2011.2108306).
\bibitem[\protect\citeauthoryear{Dudok de Wit}{2006}]{Wit}
Dudok de Wit, T.: 2006, \solphys \textbf{239}, 519 (DOI: 10.1007$/$s11207-006-0140-3).
\bibitem[\protect\citeauthoryear{Geman and Geman}{1984}]{Geman}
Geman, S., Geman, D.: 1984, {\it IEEE Trans. Pattern Anal. Mach. Intell.} \textbf{6}, 721 (DOI: 10.1109$/$TPAMI.1984.4767596).
\bibitem[\protect\citeauthoryear{Gilboa and Osher}{2007}]{Gilboa}
Gilboa, G., Osher, S.: 2007, {\it Multiscale Model. Simul.} \textbf{6}, 595 (DOI: 10.1137$/$060669358).
\bibitem[\protect\citeauthoryear{Higgins \textit{et al.}}{2011}]{Higgins}
Higgins, P.A., Gallagher, P.T., McAteer, R.T.J., Bloomfield, D.S.: 2011, \adv \textbf{47}, 2105 (DOI: 10.1016$/$j.asr.2010.06.024).
\bibitem[\protect\citeauthoryear{Humblot and Djafari}{2006}]{Humblot2006}
Humblot, F., Djafari, A.: 2006, {\it EURASIP Journal on Advances in Signal Processing}, \textbf{2006}, 1 (DOI: 10.1155$/$ASP$/$2006$/$36971).
\bibitem[\protect\citeauthoryear{Hurlburt \textit{et al.}}{2012}]{Hurlburt}
Hurlburt, N., Cheung, M., Schrijver, C., Chang, L., Freeland, S., Green, S., \textit{et al.}: 2012, \solphys \textbf{275}, 67 (DOI: 10.1007$/$s11207-010-9624-2).
\bibitem[\protect\citeauthoryear{Innes and Teriaca}{2013}]{Innes}
Innes, D.E., Teriaca, L.: 2013, \solphys \textbf{282}, 453 (DOI: 10.1007$/$s11207-012-0199-y).
\bibitem[\protect\citeauthoryear{Ireland and Young}{2009}]{Ireland}
Ireland, J., Young, C.A.: 2009, {\it Solar Image Analysis and Visualization}, Springer New York (DOI: 10.1007$/$978-0-387-98154-3).
\bibitem[\protect\citeauthoryear{Kassaye}{2013}]{Kassaye}
Kassaye, R.H.: 2013, {\it Geoinformatics $\&$ Geostatistics: An Overview}, \textbf{S1}, 12, (DOI: 10.4172$/$2327-4581.S1-012).
\bibitem[\protect\citeauthoryear{Kestener \textit{et al.}}{2010}]{Kestener}
Kestener, P., Conlon, P.A., Khalil, A., Fennell, L., McAteer, R.T.J., Gallagher, P.T., Arneodo, A.: 2010, \apj
\textbf{717}, 995 (DOI: 10.1088$/$0004-637X$/$717$/$2$/$995).
\bibitem[\protect\citeauthoryear{Kilcik \textit{et al.}}{2011}]{Kilcik}
Kilcik, A., Yurchyshyn, V.B., Abramenko, V., Goode, P.R., Gopalswamy, N., Ozguc, A., Rozelot, J.P.: 2011, \apj \textbf{727}, 44 (DOI: 10.1088$/$0004-637X$/$727$/$1$/$44).
\bibitem[\protect\citeauthoryear{Lemen \textit{et al.}}{2012}]{Lemen}
Lemen, J.R., Title, A.M., Akin, D.J., Boerner, P.F., Chou, C., Drake, J.F., \textit{et al.}: 2012, \solphys \textbf{275}, 17 (DOI: 10.1007$/$s11207-011-9776-8).
\bibitem[\protect\citeauthoryear{Bratsolis and Sigelle}{1998}]{Lucchi}
Bratsolis, E., Sigelle, M.: 1998, \aaps \textbf{131}, 371 (DOI: 10.1051$/$aas:1998274).
\bibitem[\protect\citeauthoryear{McAteer \textit{et al.}}{2005}]{McAteer}
McAteer, R.T.J., Gallagher, P.T., Ireland, J., Young, C.A.: 2005, \solphys \textbf{228}, 55 (DOI: 10.1007/s11207-005-4075-x).
\bibitem[\protect\citeauthoryear{McGrory \textit{et al.}}{2009}]{McGrory}
McGrory, C.A., Titterington, D.M., Reeves, R., Pettitt, A.N.: 2009, {\it Statistics and Computing}, \textbf{19}, 329 (DOI: 10.1007$/$s11222-008-9095-6).
\bibitem[\protect\citeauthoryear{Murphy}{2007}]{Murphy}
Murphy, K.P. :2007, {\it Conjugate Bayesian Analysis of the Gaussian Distribution}, Technical Report, University of British Columbia.
\bibitem[\protect\citeauthoryear{Narendra and Thathachar}{1974}]{Narendra}
Narendra, K.S., Thathachar, M.A.L.: 1974, {\it IEEE Trans. Syst., Man, Cybern., Syst.} \textbf{SMC-4}, 323 (DOI: 10.1109$/$TSMC.1974.5408453).
\bibitem[\protect\citeauthoryear{Petroudi, Ketsetzis, and Brady}{2004}]{PETROUDI}
Petroudi, S., Ketsetzis, G., Brady, M.: 2004, {\it WSEAS. Trans. Biology. Biomedic.} \textbf{1}, 344.
\bibitem[\protect\citeauthoryear{Priest}{2014}]{Priest}
Priest, E.R.: 2014, {\it Magnetohydrodynamics of the Sun}, Cambridge University Press (DOI: 10.1080$/$00107514).
\bibitem[\protect\citeauthoryear{Reiss \textit{et al.}}{2015}]{Reiss}
Reiss, M.A., Hofmeister, S.J., De Visscher, R., Temmer, M., Veronig, A.M., Delouille, V., \textit{et al.}: 2015, {\it Journal of Space Weather and Space Climate}, \textbf{5}, 23R (DOI: 10.1051/swsc/2015025).
\bibitem[\protect\citeauthoryear{Revathy \textit{et al.}}{2005}]{Revathy}
Revathy, K., Lekshmi, S., Nayar, S.R.P.: 2005, \solphys \textbf{228}, 43 (DOI: 10.1007$/$s11207-005-6880-7).
\bibitem[\protect\citeauthoryear{Schmieder \textit{et al.}}{2013}]{Schmieder}
Schmieder, B., Guo, Y., Moreno-Insertis, F., Aulanier, G., Yelles Chaouche, L., Nishizuka, N., \textit{et al.}: 2013, \aap \textbf{559A}, 1S (DOI: 10.1051$/$0004-6361$/$201322181).
\bibitem[\protect\citeauthoryear{Shimazaki and Shinomoto}{2007}]{Shimazaki}
Shimazaki, H., Shinomoto, S.: 2007, {\it Neural Computation}, \textbf{19}, 1503 (DOI: 10.1162$/$neco.2007.19.6.1503).
\bibitem[\protect\citeauthoryear{Tak \textit{et al.}}{2009}]{Tak}
Takeda, H., Milanfar, P., Protter, M., Elad, M.: 2009, {\it IEEE Tran. Image Process.} \textbf{18}, 1958 (DOI: 10.1109$/$TIP.2009.2023703).
\bibitem[\protect\citeauthoryear{Tan \textit{et al.}}{2010}]{Tan}
Tan, B., Zhang, Y., Tan, C., Liu, Y.: 2010, \apj \textbf{723}, 25 (DOI: 10.1088$/$0004-637X$/$723$/$1$/$25).
\bibitem[\protect\citeauthoryear{Teske and Thomas}{1969}]{Teske}
Teske, R.G., Thomas, R.J.: 1969, \solphys \textbf{8}, 348 (DOI: 10.1007$/$BF00155382).
\bibitem[\protect\citeauthoryear{Ulmschneider \textit{et al.}}{1990}]{Ulmschneider}
Ulmschneider, P., Priest, E. R., Rosner, R.: 1990, {\it Mechanisms of Chromospheric and Coronal Heating: Proceedings of the International Conference}, Springer-Verlag Berlin Heidelberg (DOI: 10.1007$/$978-3-642-87455-0).
\bibitem[\protect\citeauthoryear{Verbeeck \textit{et al.}}{2013}]{Verbeeck}
Verbeeck, C., Higgins, P.A., Colak, T., Watson, F.T., Delouille, V., Mampaey, B., Qahwaji, R.: 2013, \solphys \textbf{283}, 67 (DOI: 10.1007$/$s11207-011-9859-6).
\bibitem[\protect\citeauthoryear{Verbeeck \textit{et al.}}{2014}]{Delouille}
Verbeeck, C., Delouille, V., Mampaey, B., De Visscher, R.: 2014, \aap \textbf{561A}, 29V (DOI: 10.1051$/$0004-6361$/$201321243).
\bibitem[\protect\citeauthoryear{Werlberger, Pock, and Bischof}{2010}]{Werlberger}
Werlberger, M., Pock, T., Bischof, H.: 2010, {\it IEEE Conference on Computer Vision and Pattern Recognition} (CVPR), 2464 (DOI: 10.1109$/$CVPR.2010.5539945).
\bibitem[\protect\citeauthoryear{Wu}{1982}]{Wu}
Wu, F.Y.: 1982, {\it Rev. Mod. Phy.} \textbf{54}, 235 (DOI: 10.1103$/$RevModPhys.54.235).
\bibitem[\protect\citeauthoryear{Yoon and Kweon}{2006}]{Yoon}
Yoon, K.J., Kweon, I.S.: 2006, {\it IEEE Trans. Pattern Anal. Mach. Intell.} \textbf{28}, 650 (DOI: 10.1109$/$TPAMI.2006.70).
\end{thebibliography}
\end{document}